\documentclass[12pt,letterpaper]{JHEP3}
\pdfoutput=1
\usepackage{cite}
\usepackage{epsfig}
\usepackage{subfigure}

\graphicspath{{Figures/}}

\title{Classical Transitions in Superfluid ${}^3$He.}

\author{I-Sheng Yang$^1$\footnote{isheng.yang@gmail.com} , 
S.-H. Henry Tye$^2$\footnote{sht5@cornell.edu, iastye@ust.hk} ,  
and Benjamin Shlaer$^3$\footnote{shlaer@cosmos.phy.tufts.edu} \\
{\em 

 $^1$ ISCAP and Physics Department \\
 \hspace{0.5ex} Columbia University, New York, NY, 10027 , USA \\

 $^2$Institute for Advanced Study, \\
 \hspace{0.5ex} The Hong Kong University of Science and Technology \\
 \hspace{0.5ex} Clear Water Bay, Hong Kong,
\\  
and \\ 
 \hspace{0.5ex} Laboratory for Elementary-Particle Physics, \\
 \hspace{0.5ex} Cornell University, Ithaca, NY 14853, USA \\
 
 $^3$ Institute of Cosmology, Department of Physics and Astronomy, \\
 \hspace{0.5ex} Tufts University, Medford, MA 02155, USA 
}}

\abstract{We argue that classical transitions can be the key to explaining the long standing puzzle of the fast A-B phase transition observed in superfluid ${}^3$He while standard theory expects it to be unobservably slow.  Collisions between domain walls
are shown to be capable of reaching phases inaccessible through homogenous nucleation on the measured timescales.  We demonstrate qualitative agreements with prior observations and provide a definite, distinctive prediction that could be verified through future experiments or, perhaps, a specific analysis of existing data.}

\begin{document}

\section{Introduction}

Superfluid ${}^3$He is one of the greatest discoveries in physics\cite{Leg75,Whe75,He3}.  It provides a system free of impurities and well described by a mean field theory of the Ginsburg-Landau type.  The development of cooling technologies and detecting devices allows the superfluid phase not only to be prepared in the laboratory, but its many properties and rich phase structure can be carefully measured.  This has built a fruitful interaction between theory and the experiment.  Theoretical predictions can be frequently checked by experiments; interesting experimental results also constantly inspire the advances in the theory.

It is quite intriguing then, that one of the largest disagreements between a theoretical prediction and experimental results has been sitting in this field for over 30 years.  The first-order phase transition between the ABM phase\cite{AndMor61,AndBri73} (the A phase) and the BW phase\cite{Bal63} (the B phase) has been observed in the laboratory frequently; yet the theoretical lifetime of the A phase before the transition (due to standard thermal and/or quantum effect) is calculated to be much longer than the age of our universe.  This is the famous A-B transition puzzle.  
One may classify the possible solutions into 3 types : \\
(1) Outside influence, such as cosmic rays hitting the sample, as proposed in the baked Alaska model \cite{Leg84}; \\
(2) Quantum phenomenon, such as resonant tunneling recently proposed \cite{TyeWoh11}; \\
(3) Classical transitions. \\
Each of these explanations has its own distinctive predictions that can be explicitly checked experimentally. 

In this paper, we will not attempt to provide an overview of all proposals. It is after all a long standing puzzle in a well-studied system.  Our first step is to closely examine the status of the baked Alaska model\cite{Leg84,LegYip90}, which suggests that the transitions are triggered by incident radiation.  It is the most well-known proposal, and it is regarded by some to be the established answer already.  At the very least, the experimental data \cite{HakKru85,SchOke92,SchOsh95,SOL95,OkeBar96} seem to imply the following almost universally accepted conclusion:  {\bf The baked Alaska model plays a significant role in the A-B transition, but there can be competing/parallel effects in certain situations} \cite{SOL95}.   

However, we will point out that the same set of data admits a very different interpretation, one which is at least equally plausible.  {\bf Both radiation and ``competing/parallel effects'' are only the first step triggering the transitions;  Neither is the final control mechanism.}  Of course, this statement is not very interesting unless we can propose the final control mechanism, which we will do in this paper. 

Since the ${}^3$He system is the closest known laboratory system that mimics the cosmic landscape inspired by string theory\cite{Vil83b,GutWei83,BP} , any new discovery here may improve our understanding of the origin of our universe.  It is also possible that some of the theoretical advances in the cosmic landscape could be applied here as well \cite{TyeWoh11}.  In this paper, we will consider the effect of classical transitions\cite{BlaJos09,EasGib09,GibLam10,JohYan10}.  It is a recently realized phenomenon that first order phase transitions can happen through domain wall collisions, even if the colliding domain walls were nowhere near the new phase.  This new understanding of classical dynamics may be the key to explaining the fast A-B phase transitions in superfluid ${}^3$He.  


Based on this recent understanding, we will present the following new proposal. Because of the spin and orbital angular momentum of ${}^3$He pairs, typical lab samples of superfluid $^3$He have domains and hence intrinsic domain wall networks.  There are many ways to disturb the network, including incoming radiation and the cooling process.  However, first-order phase transitions only happen when the disturbances make domain walls collide (which can easily happen when the domains move and/or the domain walls oscillate).  The resulting field (order parameter) excursion following a collision allows the quick completion of a first-order phase transition that would have taken an exponentially long time otherwise.  This means classical transition can be the final control mechanism.  This proposal yields surprisingly good qualitative agreements with many interesting superfluid $^3$He  observations that cannot be explained by the baked Alaska model.  

Furthermore, classical transitions follow a kinematic rule of field excursions, which makes predictions very different from models relying on energetics such as the baked Alaska.  Whether a collision can induce the appropriate field excursion into the $B$ phase is controlled by the phase space structure through a single parameter, $\delta$\cite{KauKle80}, the strong coupling correction from spin fluctuations (paramagnon exchanges) \cite{BriSer74}.  With the appropriate field excursion, the transition will happen if the network disturbance is strong enough.  On the other hand, without the appropriate field excursion, disturbing it more violently will not help at all.  This makes a strong falsifiable prediction: points on the curve $\delta(P,T)= {const.}$ in the $P-T$ diagram should all together be allowed or disallowed to transit.  Our proposal will be ruled out if transitions can happen on $(P_1,T_1)$, but the same or stronger disturbances cannot trigger transitions for $(P_2,T_2)$ on the same curve.

We will present our findings in the following order.  First, in Sec.\ref{sec-he3} we give a simple review of the mean field theory description of superfluid ${}^3$He, followed by an introduction to the A-B transition puzzle.  We then will describe the baked Alaska model and examine the major positive, negative and neutral evidence for this model.  We will argue that this evidence should lead one to seek an alternative model, especially in light of the theory of classical transitions.  

In Sec.\ref{sec-CT} we review the development of classical transitions and briefly explain the dynamics.  We also made necessary generalizations in order to apply this phenomenon to superfluid $^3$He---a non-relativistic system of interacting vector fields.

Finally in Sec.\ref{sec-final}, we provide a complete description of our proposal to solve the A-B transition puzzle.  We calculate the properties relevant for classical transitions in this system, i.e., the phase space structure.  Details of this calculation are found in Appendix \ref{sec-pss}.  We list several prior observations which our proposal can explain, and provide a definite prediction. 

\section{The Helium 3 A-B Transition}
\label{sec-he3}
 
 \subsection{Introducing the Puzzle} 
 \label{sec-puzzle}

The ${ }^3$He superfluid phases are BCS states similar to conventional superconductors, except that their Cooper pairs have total spin $s=1$ and angular momentum $l=1$.  This simple difference leads to a much richer structure.

The superfluid phase is usually described by the order parameter, $\Delta_{\alpha i}$, a $3\times3$ complex matrix.  Note that it is not a tensor field, since  $\alpha$ is a spin index while $i$ describes orbital angular momentum.  They cannot directly be contracted with each other.  Spin and angular angular momentum can rotate independently, so the overall symmetry group is $SO(3)_s\times SO(3)_l\times U(1)$.

The quadratic kinetic terms are
\begin{equation}
K = \frac{1}{23T_c^2}\frac{dn}{d\varepsilon}
\left(
\partial_t\Delta^*_{\alpha i}\partial_t\Delta_{\alpha i} +
c_L^2 \partial_i\Delta^*_{\alpha i}\partial_j\Delta_{\alpha j} +
c_T^2 \varepsilon_{ijk} \partial_j\Delta^*_{\alpha k}
\varepsilon_{ilm}\partial_l\Delta_{\alpha m}
\right)~.
\end{equation}

Up to the 4th order, the potential includes all possible contractions:
\begin{eqnarray}
V &=& \frac{1}{2}\frac{dn}{d\varepsilon}\left(\frac{T}{T_c}-1\right)
\Delta^*_{\alpha i}\Delta_{\alpha i} 
+ \beta_1 \Delta^*_{\alpha i}\Delta^*_{\alpha i}\Delta_{\beta j}\Delta_{\beta j}
+ \beta_2 \Delta^*_{\alpha i}\Delta_{\alpha i}\Delta^*_{\beta j}\Delta_{\beta j}
\nonumber \\
&+& 
\beta_3 \Delta^*_{\alpha i}\Delta^*_{\beta i}\Delta_{\alpha j}\Delta_{\beta j}
+ \beta_4 \Delta^*_{\alpha i}\Delta_{\beta i}\Delta^*_{\beta j}\Delta_{\alpha j}
+ \beta_5 \Delta^*_{\alpha i}\Delta_{\beta i}\Delta_{\beta j}\Delta^*_{\alpha j}
~.
\label{eq-potential}
\end{eqnarray}

Here $dn/d\varepsilon$ is the density of states, $T_c$ is the cricitcal temperature, and $\beta_i$'s are parameters we can calcuculate from the theory.  Since $\alpha$ is the spin index, only $i$ can contract with spacetime derivatives.  The orbital angular momentum vector in $\Delta_{\alpha i}$ is just like a vector field, and has transverse and longitudinal velocities $c_T$ and $c_L$, respectively.

Above $T_c$, the system is in the normal Fermi liquid state with $\langle \Delta_{\alpha i}\rangle=0$.  For $T<T_c$, the symmetry is spontaneously broken, similar to behavior in a standard $(-m^2\phi^2+\lambda\phi^4)$ potential.  Here the rich symmetry of $SO(3)_s\times SO(3)_l\times U(1)$ allows two different types of condensation.
One of them is the A-phase\cite{AndMor61,AndBri73}, usually described by the matrix
\begin{equation}
\Delta(A) = \Delta_A 
\left( {\begin{array}{ccc}
 1 & i & 0  \\
 0 & 0 & 0  \\
 0 & 0 & 0  \\
 \end{array} } \right)~.
\label{eq-A}
\end{equation}
We can see that an $SO(3)_l$ rotation along the $z$ axis is degenerate with a complex phase.

The other condensate type is the B-phase\cite{Bal63}, usually denoted
\begin{equation}
\Delta(B) = \Delta_B 
\left( {\begin{array}{ccc}
 1 & 0 & 0  \\
 0 & 1 & 0  \\
 0 & 0 & 1  \\
 \end{array} } \right)~.
\label{eq-B}
\end{equation}
A simultaneous rotation on both $SO(3)$'s leaves this ``isotropic" phase invariant.

During the cooling process, the instability into the A-phase actually occurs first.  Its free energy also remains lower than that of the B-phase down to some $T_{AB}<T_c$.  There is no way to execute a ``super fast'' cooling to bypass the above temperature range where the A-phase is dynamically preferred.  Therefore, the superfluid phase transition always goes to the A-phase first in all laboratory experiments.

Below $T_{AB}$, the A-phase becomes metastable, the B-phase becomes the true vacuum, and they are separated by a free energy barrier.  The superfluid may then undergo a first-order phase transition---nucleating a critical bubble of the B-phase that grows to convert the entire sample.  The tension $\sigma$ and the free energy difference $\delta V$ between two phases are not only theoretically calculated \cite{KauKle80,Schopohl}, but also experimentally measured to a good accuracy\cite{OshCro77}.  We can use these parameters to calculate the critical radius $R_c$ and the tunneling rate $\Gamma$. \footnote{The (2 or 3) is the difference between a thermal instanton and a quantum instanton.  Their rates are both unobservably small and we are quoting the relatively higher one, i.e., the one for zero temperature \cite{BL}.}
\begin{eqnarray}
R_c &=& (2{\rm \ or\ }3)\frac{\sigma}{\delta V}\sim1\mu {\rm m}~, \nonumber \\
\Gamma &\sim& \exp\left[-S_{\rm instanton}\right]\lesssim10^{-20000}~. 
\end{eqnarray}
Although a critical bubble can be easily contained in a lab sample, the lifetime before this decay, regardless of the dimensionful prefactor in $\Gamma$, is longer than the age of the universe.

Strangely, the A-B phase transition almost always is found to happen when the sample is supercooled significantly below $T_{AB}$.  It happens in a few minutes or a few hours.  In a different system, this kind of discrepancy might be explained by impurities or boundary effects.  However ${ }^3$He superfluid has no impurities.  Only ${ }^4$He can dissolve it in and can be controlled to less than one molecule per mole in the sample.   The boundary effect actually stabilizes the A-phase\cite{FreGer88}, as the boundary defines a direction, while the B-phase is isotropic\footnote{Recently, it was argued that the boundary effect, although favoring the A-phase over the B-phase, cannot be totally excluded from the discussion\cite{BalMiz00}.}.

This is the famous A-B transition puzzle in superfluid ${ }^3$He.

 \subsection{baked Alaska Model}
 \label{sec-BAM}
One exotic, but seemingly viable explanation after eliminations, is the baked Alaska model\cite{Leg84}.  It was proposed that the transition is facilitated by cosmic rays.  As highly boosted muons pass through the sample, secondary electrons are produced.  Those that stop do so very abruptly, and thereby act as point sources to locally heat up the system.

Note that this cannot be thought of as a conventional thermal activation.  Thermodynamically, a hot spot cools down by diffusion and the cooling rate is controlled by a thermal scale.  As mentioned earlier, this process most likely leads to the A-phase again. Additionally, during diffusion the exterior is cooler than the interior and so always condenses first to whichever phase it is exposed to, which is A-phase.

Fortunately, as pointed out in\cite{Leg84}, the energy released by an electron goes to quasiparticles with mean free path $\sim2\mu {\rm m}$, which is $\gtrsim R_c$.  Therefore, for physics at the scale $R_c$, it is not a conventional thermal system.  The best description in this situation is (still) unknown, so Leggett made a {\it conjecture}.

The baked Alaska conjecture is the following.  Since most of the energy is carried by quasiparticles with mean free path around $2\mu {\rm m}$, below such length scale, there is no diffusion and it is appropriate to picture the free motion of quasiparticles as a hot and expanding shell.  The purpose of this hot shell is twofold.  It heats up and then shields the interior from the ambient A-phase region to allow an independent cooling. Also, regions swept through by this hot shell effectively go through a rapid cooling, which avoids the usual cooling constraint and has a chance to settle into the B-phase.  Since $R_c<2\mu {\rm m}$, it is possible to have regions in the B-phase large enough to expand even after the hot shell diffuses away.

Due to the lack of viable alternatives, this proposal has remained the center of focus for over 30 years.  Before moving on to examine the evidence, we would like to emphasize some conceptual hurdles faced by this portion of the model.   It describes regions undergoing ``heating'' and ``cooling''  on scales below the thermalization scale.  Namely, it applies thermal intuition on potentially non-thermal dynamics.  There might be situations where it is a good description, but some skepticism is warranted.  Additionally, in comparison with data described in Sec.\ref{sec-data}, it is a problematic portion of the baked Alaska model.

As an interesting sidenote, in Sec.\ref{sec-CT} we will mention that during the development of the classical transition theory, a similar conjecture was made and later proved to be insufficient to provide the whole answer.  Together with the experimental data, it is a sign that the baked Alaska model could at least be improved.

 \subsection{Examining the Evidence}
 \label{sec-data}

Our analysis here closely follows \cite{SOL95} which analyzed old data from A-B phase transition experiments.  It also summarizes the results from a series of more recent experiments at Stanford\cite{SchOke92,SchOsh95} which were designed to test the baked Alaska model. \\
\ \\ 
There are two major positive pieces of evidence.
\begin{itemize}
\item{\bf P1.} The nucleation rate becomes significantly higher if the samples are exposed to a radiation source.  The difference is reasonably proportional to the strength of radiations.
\item{\bf P2.} The temperature dependence is the same (up to an overall scaling) with or without the presence of the radiation source.
\end{itemize}
\ \\
There are also three major negative pieces of evidence.
\begin{itemize}
\item{\bf N1.} The theoretical prediction has $\ln\Gamma\propto -R_c^n$ with $3<n<5$\cite{LegYip90}, but the best fit to data is $n=3/2$, and is unlikely to be $n>2$.
\item{\bf N2.} The measured temperature dependence stays the same when the dominant incoming radiation is switched from electrons to neutrons.  Theoretically, there should be a difference.
\end{itemize}
\ \\
Finally, there is some evidence pointing toward competing/parallel mechanisms. In many occasions people found transitions uncorrelated with incoming radiation.
\begin{itemize}
\item{\bf C1.}  A direct search of cosmic ray effect detected no correlation between a cosmic ray event and the A-B transition in a superfluid He-3 sample \cite{Swift}.
\item{\bf C2.}  Transition rates are significantly higher in containers with boundary roughness (on the length scale $\mu {\rm m}$).
\item{\bf C3.}  Different cooling rates can increase or decrease the transition rate under different circumstances.
\item{\bf C4.}  Mechanical disturbances usually cannot trigger the transition, except for samples with Si wafers (introduced to create surface roughness) on the brink of a transition\cite{OkeBar96}.
\item{\bf C5.}  At strong magnetic field and low temperature, the magnetic field seems to only affect the rate through its effect on the free energy difference $\Delta V$.  However in high $T$/low $H$, it behaves differently.
\end{itemize}

Indeed (P1) establishes a solid correlation between radiation and the phase transitions.  The easiest interpretation of (P2) is that without adding a radiation source, the transitions are triggered by the background radiations through the same baked Alaska mechanism.  However, even if we only consider positive evidence, it does not exclude the possibility that the background trigger is not due to radiation, as long as there is a final control mechanism providing the unique temperature dependence.

Furthermore, (N1) and (N2) clearly suggest that the baked Alaska model requires some modifications.  The quantitative prediction that disagrees with data comes directly from the baked Alaska conjecture.  Together with the theoretical concern mentioned in Sec.\ref{sec-BAM}, it is reasonable to seek out alternatives, perhaps in the form of this final control mechanism.  In fact, according to  \cite{SOL95}, {\it ``While one might argue that radiation could be causing nucleation through some other mechanism, it is difficult to imagine another and none has been proposed in the 18 years that this problem has been outstanding.''}, the lack of alternatives seems to be one of the main reasons to keep the baked Alaska model in place. 

Thanks to the discovery classical transitions, we do have such an alternative.  We believe it is worth exploring, especially if it can also explain some of (C1) to (C5).  We will argue that is indeed the case.

\section{Classical Transitions}
\label{sec-CT}

\begin{figure*}[ht]
   \centering
   \includegraphics[width= 2.5in]{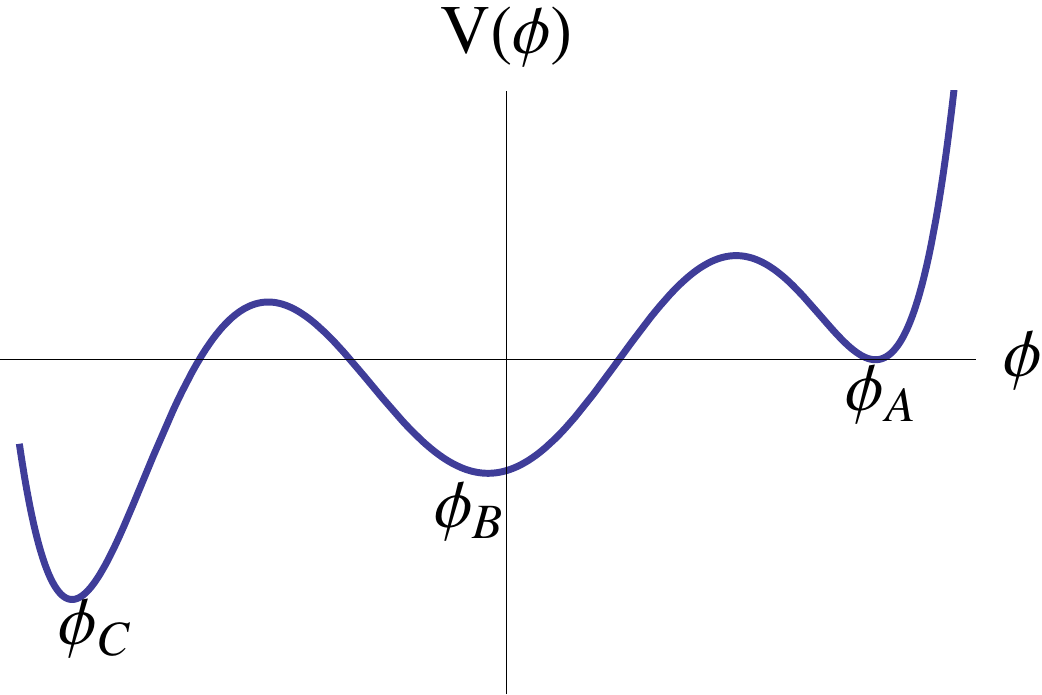} 
   \caption{\label{fig-V} A typical potential allowing classical transitions.}
\end{figure*}

\begin{figure*}[ht]
   \centering
   \includegraphics[width= 5in]{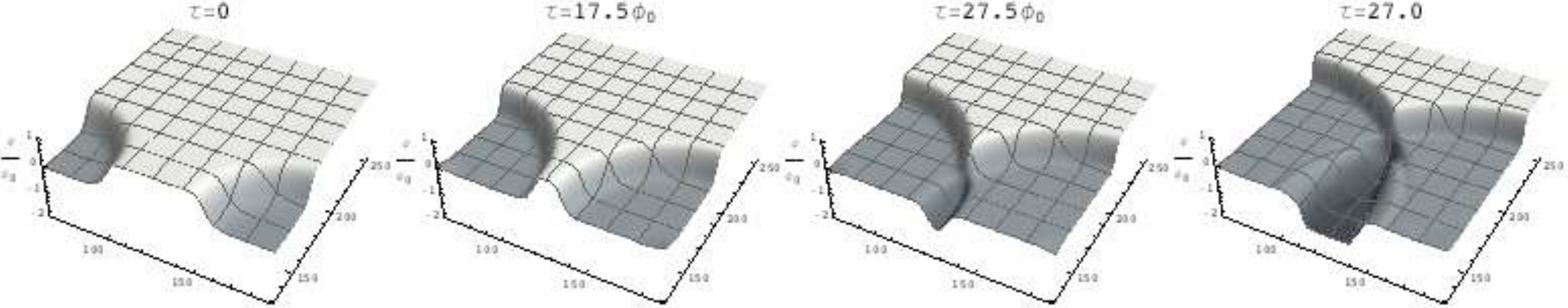} 
   \caption{\label{fig-CT} Lattice simulation showing a classical transition.  The bulk value is false vacuum $\phi_A$.  Initially there were two bubbles of $\phi_B$.  Their collision results in a transition to $\phi_C$.}
\end{figure*}

Classical transition is a way to achieve the result of a first-order phase transition through purely classical behaviors.  Consider a simple Lagrangian as in\cite{EasGib09},
\begin{equation}
{\cal L} = \frac{1}{2}(\partial\phi)^2 - V(\phi)~,
\end{equation}
where $V$ is a potential with three local minima, as shown in Fig.~\ref{fig-V}.  It was observed in high resolution lattice simulations that when two domain walls between vacua A and B collide, the collision region can sometimes go to vacuum $C$ as in Fig.~\ref{fig-CT}.

Note that although we started by slamming two domain walls together instead of being hit by a quantum of radiation, the second half of the induced phase transition is very similar to the baked Alaska model.  We generate two energetic ``shells'' to protect the middle region, give it a chance to settle into a different minimum, and guarantee its subsequence growth.

Not surprisingly, a very similar conjecture was made to explain this phenomenon.  Originally in\cite{EasGib09} this was described as an ``energetic'' effect.  The highly boosted domain walls provide enough energy during the collision to push the field $\phi$ both above the potential barrier and toward a field excursion.  This predicted that higher energy can overcome a higher barrier and reach a $\phi_C$ further away.

The prediction was soon realized to be incomplete.  In\cite{GibLam10} it was shown that a threshold energy is needed to cross the barrier, but adding more energy does not push the field further.  The transition only happens when $(\phi_C-\phi_B)\lesssim(\phi_B-\phi_A)$.  

We think there is a general lesson learned here, which should also apply to the baked Alaska conjecture.  The collision happens in a very short time scale that our intuition of ``available energy'' does not work, just as the intuition of ``cooling'' does not work in a non-thermal system.  A microscopic description is needed, and in this case it is the kinematics of the field that controls the transition.

 \subsection{The Mechanism}

\label{sec-CTm}

The classical transition behavior comes from the equation of motion.
\begin{equation}
\frac{\partial^2\phi}{\partial t^2}-\frac{\partial^2\phi}{\partial x^2}
=-\frac{\partial V(\phi)}{\partial \phi}~.
\label{eq-eom}
\end{equation}

Given two degenerate minima in $V$ and a barrier between them, we can find a static interpolating solution $f(x)$ that solves
\begin{equation}
-f''(x)=-\frac{\partial V(f)}{\partial f}~,
\end{equation}
with $f(-\infty)=\phi_L$ and $f(\infty)=\phi_R$.  It looks like a domain wall in the middle where the field obviously changes, and stays exponentially close to either vacuum on each side.  We can arbitrarily boost this solution, $f(\frac{x-vt}{\sqrt{1-v^2}})$ will also be a solution of Eq.~(\ref{eq-eom}).

When the vacua are not degenerate, the domain wall will feel a pressure difference and a constant acceleration.  In the thin wall limit, at every instant the accelerating solution is approximated by a boosted solution.  

The behavior of Eq.~(\ref{eq-eom}) in these highly boosted solutions is crucial to understanding classical transitions.  In the vacuum region it is trivially $0=0$.  Within a highly boosted domain wall, both terms on the left hand side are enhanced by a large factor, $\gamma^2=1/(1-v^2)$.  So the zeroth order solution is to ensure they cancel each other,
\begin{equation}
\frac{\partial^2\phi}{\partial t^2}-\frac{\partial^2\phi}{\partial x^2}\sim0~,
\label{eq-free}
\end{equation}
which is a free field equation.  The leftover of this approximate cancellation have to agree with $-\frac{\partial V(\phi)}{\partial \phi}$, but that is akin to next order in perturbation theory.  The expansion parameter is $1/\gamma^2$.

Now consider three regions in three different vacua, with two highly boosted domain walls moving toward each other.  Before they collide, the solution looks like
\begin{eqnarray}
\phi(x,t)&=&\phi_M + f_L\left(\frac{-x+vt}{\sqrt{1-v^2}}\right)+
f_R\left(\frac{x+vt}{\sqrt{1-v^2}}\right) 
\nonumber \\
&\approx&
\phi_M + (\phi_L-\phi_M)\Theta(-x+t)+(\phi_R-\phi_M)\Theta(x+t)~,
\end{eqnarray}
where $\phi_L$, $\phi_M$, $\phi_R$ are the vacuum field values of the left, middle, right regions; $f_L$ and $f_R$ are the domain wall solutions from the middle vacuum to left and right vacua, which are almost step functions when highly boosted.  If all we have to solve is a free field equation, Eq.~(\ref{eq-free}), then the middle region initially at
\begin{equation}
\phi(0,t<0)=\phi_M
\end{equation}
just goes to
\begin{equation}
\phi(0,t>0)=\phi_M + (\phi_L-\phi_M) + (\phi_R-\phi_M) 
=\phi_L+\phi_R-\phi_M~. 
\label{eq-CT}
\end{equation}
This is why adding more energy will not help at all.  A fixed ``field excursion'' is determined by microscopic kinematics.  If this field excursion is near a vacuum, then Eq.~(\ref{eq-eom}) continues to be solved trivially, and one finds a classical transition into this vacuum.  This result directly generalizes to multiple fields with standard kinetic terms,
\begin{equation}
\frac{\partial^2\vec{\phi}}{\partial t^2}
-\frac{\partial^2\vec{\phi}}{\partial x^2}
=-\vec{\nabla}V~,
\end{equation}
for which
\begin{eqnarray}
\vec{\phi}(0,t<0)&=&\vec{\phi}_M~, \nonumber \\
\vec{\phi}(0,t>0)&=&\vec{\phi}_L+\vec{\phi}_R-\vec{\phi}_M~. 
\end{eqnarray}

\subsection{Basin of Attraction}
\label{sec-boa}

Eq.~(\ref{eq-CT}) provides a field value right after the collision.  For the evolution afterward, we can no longer ignore the potential unless it is exactly another local minimum.  Usually we vaguely say if the field excursion enters the ``basin of attraction'' of some minimum, it will eventually roll there.

\begin{figure*}[ht]
   \centering
   \includegraphics[width= 3.5in]{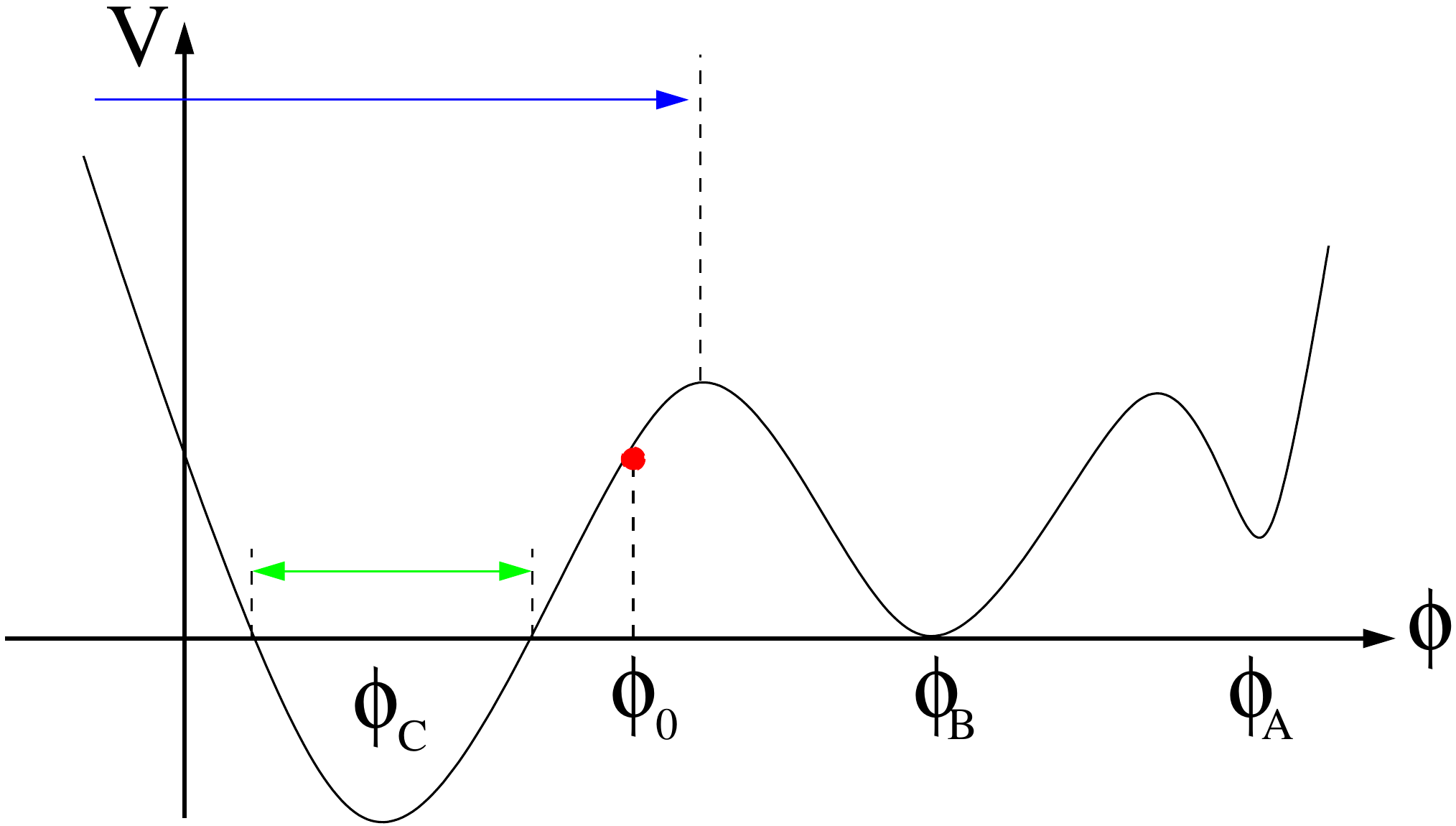} 
   \caption{\label{fig-boa} Showing the na\"ive basin of attraction spanned by the blue arrow, and the absolute basin of attraction spanned by the green arrow.}
\end{figure*}

\begin{figure*}[ht]
   \centering
   \includegraphics[width= 1.2in]{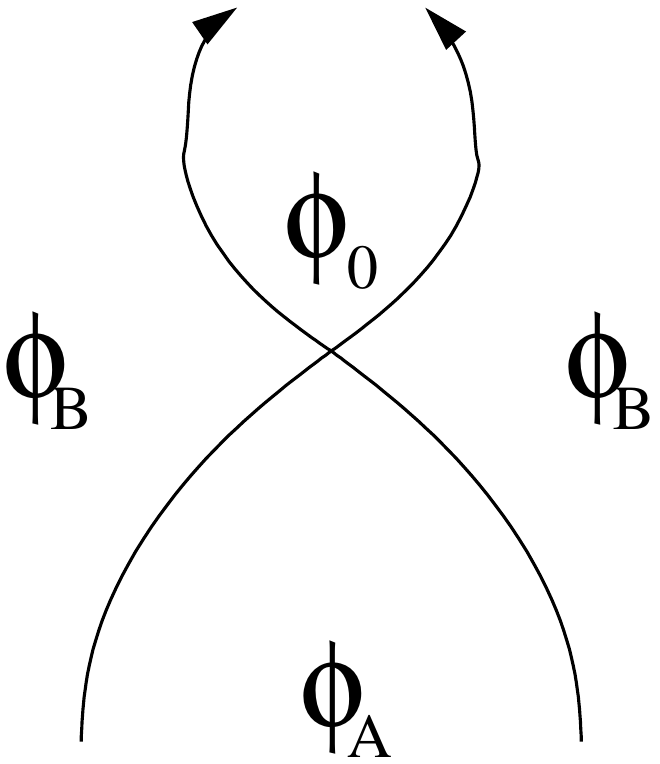} 
   \caption{\label{fig-comp} The field interpolation between $\phi_0$ and $\phi_B$ goes over a barrier, which behaves like a quasi-domain-wall.  The pressure difference makes them turn around and may collapse the middle region before the field rolls from $\phi_0$ to $\phi_C$.}
\end{figure*}

This is somewhat confusing.  As in Fig.~\ref{fig-boa}, one might think we just have to go over the top of the energy barrier between $\phi_B$ and $\phi_C$, for example to $\phi_0$.  It is not this simple.  As $\phi$ rolls down, there is a competing effect as shown in Fig.~\ref{fig-comp}.  Since $V(\phi_0)>V(\phi_B)$, the middle region will start to give in to the surrounding region in $\phi_B$.  It may recollapse entirely before reaching $\phi_C$.

On the other hand, this is the only competing effect.  So if $V(\phi_0)<V(\phi_B)$, then it will always roll down to $\phi_C$.  This allows us to define two regions: the na\"ive basin of attraction (just cross the barrier) and the absolute basin of attraction, $V<V(\phi_B)$.  The real basin attraction which defines whether a classical transition can happen or not, must be somewhere between these two. As the temperature below $T_{AB}$ decreases towards zero, the size of the actual basin of attraction is expected to grow.

\subsection{Sound Speeds}

\label{sec-CTcs}

In Sec.\ref{sec-CTm} we used the term ``highly boosted'', which may seem only applicable in relativistic systems.  That is not true.  Adding a sound speed to Eq.~(\ref{eq-eom}),
\begin{equation}
\frac{\partial^2\phi}{\partial t^2}-c_s^2\frac{\partial^2\phi}{\partial x^2}
=-\frac{\partial V(\phi)}{\partial \phi}~,
\end{equation}
is equivalent to rescaling the $x$ coordinate.  Alternatively, we can define $\gamma = (1-v^2/c_s^2)^{-1/2}$, the entire arguement goes through the same way.  The domain wall becomes highly boosted when it approaches the sound speed.  

The real problem is, in condensed matter systems, we can have multiple sound speeds.  For example in a system with a vector field, there are the transverse modes and the longitudinal mode, with $\sqrt{3}c_T=c_L$.
\begin{eqnarray}
\frac{\partial^2\phi}{\partial t^2}-c_T^2\frac{\partial^2\phi}{\partial x^2}
&=&-\frac{\partial V(\phi,\psi)}{\partial \phi}~, \nonumber \\
\frac{\partial^2\psi}{\partial t^2}-c_L^2\frac{\partial^2\psi}{\partial x^2}
&=&-\frac{\partial V(\phi,\psi)}{\partial \psi}~.
\label{eq-2sound}
\end{eqnarray}
Now what happens to domain walls in this system?

First of all, the static domain walls are easy.
\begin{eqnarray}
-\frac{\partial^2(c_T\phi)}{\partial x^2}
&=&-\frac{\partial V(\phi,\psi)}{\partial (c_T\phi)}~, \nonumber \\
-\frac{\partial^2(c_L\psi)}{\partial x^2}
&=&-\frac{\partial V(\phi,\psi)}{\partial (c_L\psi)}~.
\end{eqnarray}
Simply rescale the sound speeds into the fields definition, $\tilde{\phi}=c_T\phi$, $\tilde{\psi}=c_L\psi$, and it is equivalent to a standard multifield problem\footnote{Note that this means the structure (therefore the tension) of a domain wall depends on its orientation.  It is worth further study but quite unlikely to dramatically change the exponentially small transition rate.}.

\begin{figure*}[ht]
\begin{center}
\subfigure[]
{\includegraphics[width=.45\textwidth]{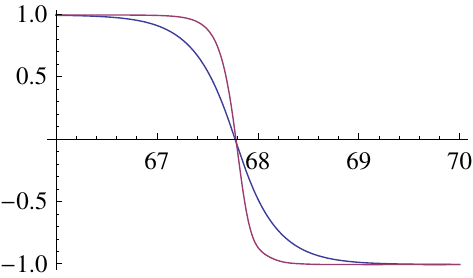}}
\hspace{1in}
\subfigure[]
{\includegraphics[width=.30\textwidth]{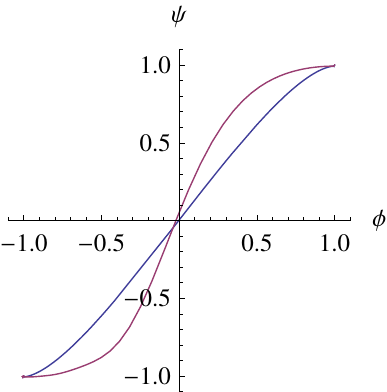}}
\caption{A domain wall between two vacua at (1,1) and (-1,-1).  The left figure is the field profile of a moving domain wall.  The red curve is the transverse mode $\phi$, which is narrower (more boosted) than the longitudinal mode $\psi$ (blue curve).  The right figure is the domain wall trajectory in the field space.  The blue path is the same moving domain wall as in the left figure.  The red path is the stationary domain wall.  When $v\rightarrow c_T$, the path will become a step function.  The $\phi$ interpolation becomes very narrow but $\psi$ does not.
\label{fig-path}}
\end{center}
\end{figure*}

For a domain wall moving in a constant velocity, we can put in the anzatz $\phi(x-vt)$, $\psi(x-vt)$ into Eq.~(\ref{eq-2sound}) and get
\begin{eqnarray}
(v^2-c_T^2)\phi''
&=&-\frac{\partial V(\phi,\psi)}{\partial \phi}~, \nonumber \\
(v^2-c_L^2)\psi''
&=&-\frac{\partial V(\phi,\psi)}{\partial \psi}~.
\label{eq-2soundv}
\end{eqnarray}
For $v^2<c_T^2$, we can similarly rescale by $\tilde{\phi}=\sqrt{c_T^2-v^2}\phi$, $\tilde{\psi}=\sqrt{c_L^2-v^2}\psi$, and get to a standard multifield problem.  But note that the potential seen by the rescaled fields depends on the velocity, and therefore the path in the field space $(\phi,\psi)$ depends on the velocity, as shown in Fig.\ref{fig-path}.  For example in the limit $v^2\rightarrow c_T^2$, the equation of motion for $\phi$ is just $\frac{\partial V}{\partial\phi}=0$.  This means we should globally minimize $V$ in the $\phi$ direction even if it leads to a discontinuity.

How about $v^2>c_T^2$?  First of all, if the domain wall starts to accelerate from zero velocity, the gradient energy will diverge as it approaches $c_T$.  Even if we try to setup an initial condition with something faster than $c_T$, Eq.~(\ref{eq-2soundv}) implies an instability in the small perturbations of $\phi$.  When the domain wall moves faster than $c_T$, it emits perturbations traveling at $c_T$. These are no different from Cherenkov radiation!  They pose a large energy cost, causing the domain wall to quickly slow below $c_T$.  As confirmed by simulations\footnote{We have only done simple $1+1$D simultions in Mathematica, which is sufficient for our purpose.  The vector field system itself might worth further study with higher resolution simulations.}, this sound speed barrier keeps the velocity of an accelerating domain wall at $c_T$.  Only the transverse mode $\phi$ in the domain wall profile becomes highly boosted like a step function.  The effective boost for the longitudinal mode $\psi$ remains of order one, $\gamma = (1-c_T^2/c_L^2)^{-1/2}=\sqrt{3/2}$.

Although there are very special potentials (for example Sine-Gordon) where classical transitions happen at an arbitrarily low boost, in general $\gamma=\sqrt{3/2}$ is not enough.  Here we have a rather peculiar situation---a half classical transition.  Indeed $\phi$ classically transits but $\psi$ does not.
\begin{eqnarray}
\phi(0,t>0)&=&\phi_L+\phi_R-\phi_M~, \\
\psi(0,t>0)&=&\rho\left[(\psi_L-\psi_M)+(\psi_R-\psi_M)\right]+\psi_M 
\\ \nonumber &=&\rho(\psi_L+\psi_R) + (1-2\rho)\psi_M~.
\label{eq-CTcs}
\end{eqnarray}
where $\rho$ is a factor between $0$ and $1$ which depends on various details and can fortunately be neglected in this paper.

Note that this analysis is valid for a $(1+1)$ dimensional collision, namely a head-on collision.  In a relativistic system we can always boost to this frame, but in a condensed matter system we cannot.  We will then assume that there will occur nearly head-on collisions in the system for which our analysis is valid, and this can be taken as an approximation (or an upper bound) for non-head-on collisions.\footnote{Actually, part of the results in Fig.~\ref{fig-CT} are circumstantial evidence showing that non-head-on collisions lead to the same transition.}

\section{Proposing a New Answer to the A-B Transition Puzzle}
\label{sec-final}

Having established that classical transitions are part of the field dynamics in this system, we are ready for a new proposal to explain the A-B phase transition puzzle.  Since we only use the standard equations of motion, our proposal does not involve any novel (unconfirmed) dynamics.  We will show how the superfluid sample allows domain wall collisions to take place frequently.  Then we show these collisions can lead to classical transitions.

 \subsection{A Dynamical Domain Wall Network}
 \label{sec-break}

The A-phase $^3$He superfluid is a nearly ferromagnetic system.  While cooling from the normal phase over a time scale $\tau\sim10^3 - 10^4s$\cite{ParRuu95}, the second-order transition into the A-phase has the initial coherence length $\xi\sim10^{-2}{\rm cm}$\cite{KibVol97}.  This is the typical size of regions, though all in the A-phase degenerate valley, having their order parameter $\Delta_{\alpha i}$ pointing in different directions.  These regions are separated from each other by domain walls whose thickness is the ``healing length'' from the dipole alignment effect, $\xi^A_D\sim8\mu {\rm m}$.\cite{He3}

One may worry that unlike a solid magnet, for which the spins are frozen after settling into these regions, the fluid nature can erase such structure.  For example, if the degeneracy in the A-phase is further broken, the region with the lowest free energy might expand and eliminate the others.  We think that is unlikely for two reasons.

The strongest breaking effects come from the container boundary and external magnetic fields.  Near the boundary, the orbital angular momentum prefers to align normally to the surface, and different parts of the boundary will not agree with each other.  In particular, if the surface roughness has a length scale between $\xi$ and $\xi^A_D$, it will actually pin down the domain wall structure near the boundary.  In the middle region away from the boundaries, the magnetic moment prefers to align with the external magnetic field (if it is strong enough).  Note that these are vector effects, so there are still degeneracies remaining.  There is really no single orientation within the A-phase that dominates others through out the sample.

Second, there is an interesting hierarchy in this domain wall network.  The domain walls intersect each other in string-like and point-like defects (vortices).  Unlike the domain walls which are sustained by the small bending energy along the A-phase valley, these lower dimensional defects are generally forced to go through $\Delta_{\alpha i}=0$, which is a peak in the potential.  In other words, they are much heavier.  In the time scale that the domain walls are dynamical, we can treat the strings and vortices as stationary.    While the domain walls may bend responding to further degeneracy breaking, the strings and the vortices might be enough to pin down the overall network structure.

\begin{figure*}[ht]
\begin{center}
\includegraphics[width=8cm]{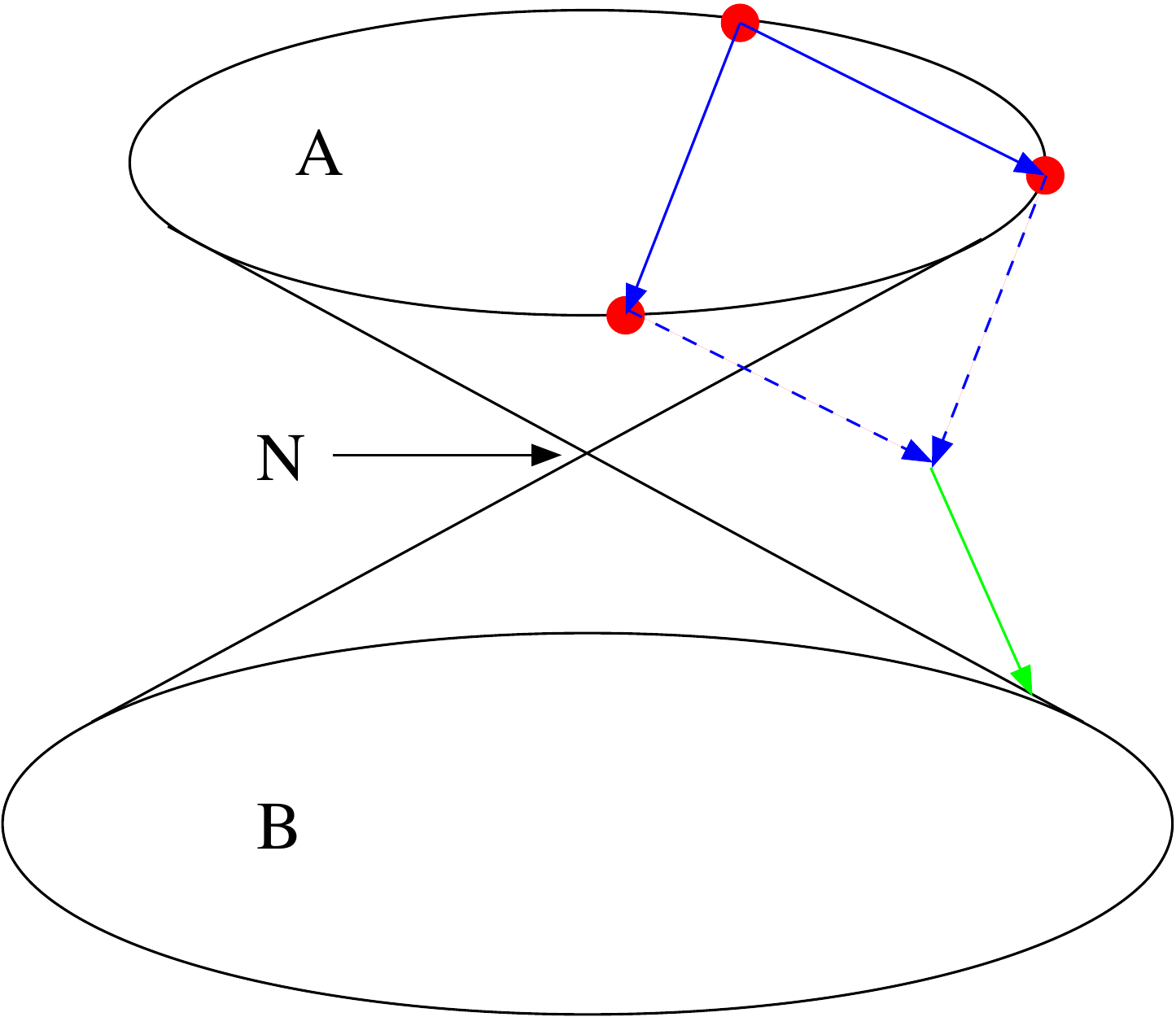}
\caption{The double cone is a cartoon picture of the superfluid ${}^3$He phase space structure.  The middle apex is the symmetric normal phase $N$.  The top and the bottom circles are the spontaneous broken phases A and B.  Three sub-phases of A (red dots) involved in a domain wall collision can lead to a field excursion as the blue arrows demonstrated.  This field excursion can roll down to the B-phase and complete an A-B transition.
\label{fig-CTAB}}
\end{center}
\end{figure*}

Therefore, a lab sample is never homogeneously in one particular configuration of the A-phase.  It consists of many  sub-domains of various sub-phases of A, randomly formed during the second-order transition from the normal phase.  The domain walls between them stay in an equilibrium configuration when the sample is not disturbed.  Once cooling has started, or responding to incoming radiation, the system will be temporarily out of equilibrium, and domain walls can move to collide with each other\footnote{One can also imagine nucleation of sub-phase bubbles, which will expand and eventually collide with existing domain walls.  The exponentially small tunneling rate in Sec.\ref{sec-he3} is only for an A-B transition.  The transition rate between sub-phases in A has not been estimated and need not be small.  We prefer to use existing domain walls as our core mechanism for it matches past observations better.}.  The stochastic behavior comes from the repeated collisions between different combinations of random sub-phases, as depicted in Fig.\ref{fig-CTAB}.  Each of these collisions has a chance to induce a field excursion into the B-phase basin of attraction.

Superfuild $^3$He is a complicated system.  There might be more than one dominant effect disturbing the domain wall network\footnote{We believe that in experiments with very smooth boundaries\cite{SchOke92,SchOsh95}, radiation has the dominant effect.  In \cite{HakKru85}, we believe the cooling process and the boundary roughness are the main driving forces for domain wall motions.}, and the detailed dynamics of a disturbed network are not easy to picture\footnote{For example, the strings and vortices, though relatively stationary, can still annihilate with each other, and might suddenly ``pop'' when the surrounding domain walls accumulate a strong stress on them.  This will subsequently result in further domain wall motion.}.  Fortunately, classical transitions may allow us to ignore these details.  As long as many collisions happen above the energy threshold, the final control mechanism will be the field excursions.

 \subsection{Criteria for Classical Transitions}
 \label{sec-crit}

As mentioned in Sec.\ref{sec-CT}, a domain wall collision needs to pass a certain energy threshold and induce an appropriate field excursion to complete a classical transition.  Here we will show that both conditions are satisfied in superfluid ${}^3$He.

The simplest way to think about the energy threshold was discussed in \cite{JohYan10}.  We can just look at Fig.\ref{fig-comp} and imagine the collision as a scattering problem of domain walls.  Two incoming domain walls between sub-phases of A need to become two outgoing domain walls between phase A and phase B\footnote{This is particularly useful when the incoming domain walls are lighter than the outgoing ones, since it guarantees a large boost.  For the opposite case it will be more subtle.  Fortunately our system is the easier case.}.  $\sigma_{AB}$ comes directly from an energy barrier in the mean field theory.  The same mean field theory tells you that any 2 sub-phases of A are connected in a flat direction.  Thus, their rest tensions usually satisfy the hierarchy
\begin{equation}
\sigma_{AB}\gg\sigma_{AA}~.
\end{equation}
Therefore, the incoming domain walls need a large kinetic energy to make up the difference.

The kinetic energy comes from the potential difference between sub-phases, $V_{ij}\equiv|V(A_i)-V(A_j)|$.  Their precise values are beyond our current knowledge, but since boundary effects can totally suppress the A-B transition\cite{FreGer88}, some $V_{ij}$ is comparable to $V_{AB}\equiv[V(A)-V(B)]$.  In that case, the length scale we need to accelerate the domain wall so that its kinetic energy approaches $\sigma_{AB}$, is roughly the critical bubble size $R_c\sim 1\mu {\rm m}$.  The typical size of a domain in one sub-phase of A is $\xi\sim10^{-2}{\rm cm}$ as mentioned earlier.  This means we do have enough kinetic energy down to $V_{ij}\sim V_{AB}/100$.   It is quite plausible for many collisions to pass the energy threshold.

As for the field excursions, we put the detailed analysis in Appendix \ref{sec-pss} and only summarize the relevant results here.  Sub-phase collisions in A never take the order parameter $\Delta_{\alpha i}$ directly to the B-phase. This is a desired property.  Unlike a nucleated bubble, a B-phase region created by a classical transition will be nearly planar and surrounded by domain walls still moving apart. Most likely it will prevail as long as $T<T_{AB}$.  But experimentally the transition is known to only occur at an even lower temperature threshold $T\sim0.7T_{AB}$.  So if a collision directly created a B region, our proposal would have already been ruled out.

Using the behavior of the na\"ive and the absolution basins of attraction, Appendix \ref{sec-pss} also shows that the field excursion starts to go into the B-phase basin of attraction at some finite temperature, and moves in deeper at lower temperatures.  This means classical transitions is a possible candidate to solve the A-B transition puzzle.

 \subsection{Agreement with Past Experiments}
 \label{sec-exp}
 
Qualitatively, this new proposal of domain wall network classical transitions can agree with numerous strange observations made in past experiments, even those ones seemingly contradicting each other.  Many of them, if we stick with the baked Alaska model, would have to be explained by competing/parallel mechanisms.

\begin{itemize}
\item {\it Transitions are induced by radiations, but do not agree with the explicit dependence given by the baked Alaska model.}  Because incoming radiation only disturb the equilibrium and allows further domain wall collisions.
\item {\it Transitions start at $T\sim0.7T_{AB}$ instead of $T_{AB}$.}  This could be the temperature for which the possible field excursions start to enter the B basin of attraction.
\item {\it Transitions happen only during cooling but not heating.}  While bringing a supercooled sample back to a higher temperature, the induced domain wall motion is almost a reverse process of that during cooling.  If the involved sub-phases collisions did not lead to the B-phase basin of attraction during cooling, neither will they during heating.  This explains why temporarily heating a tube segment in the Stanford experiments failed to induce a transition.  This also explains why one particular sample in \cite{HakKru85} did not go to the B-phase at all.  Because the random A sub-phases in its domain wall network failed to contain a correct combination.  Within the temperature range they tried, none of the domain wall collisions can induce an appropriate field excursion to go to the B-phase. 
\item {\it The A-phase is more stable with a smoother container boundary.}  Because a rougher boundary encourages more sub-domains, and may result in nonuniform cooling.  Both effects increase the chance of collisions between different sub-phases.
\item{\it Transitions are generally not induced by mechanical disturbances nor high resonant RF.}  Bulk disturbances shift the entire system instead of affecting the structure of the domain wall network.  To possibly do that, the disturbance needs a wavelength comparable to the typical size of a sub-domain $\sim 10^{-2}{\rm cm}$.  The frequency needs to be $\sim 5\times10^5$Hz for sound waves and $\sim3\times10^{12}$Hz for EM waves.
\item{\it As an exception to the previous observation, samples with Si wafers on the brink of transitions does immediately go to the B-phase after a gentle tap.} \cite{OkeBar96}  Si wafers have surface roughness which pins down some of the domain wall network.  So the relative motion between the Si wafers and the bulk sample does disturb the network equilibrium.
\item {\it Fast cooling from $T<T_c$ allows the A-phase to reach a lower temperature in the Stanford experiments.}  A shorter process allows less combinations of A sub-phases collisions to be tried.  Note that fast cooling may allow the domain walls to accelerate more as it disturbs the equilibrium faster.  However, as long as most of the collisions happen above the threshold energy, exploring field space excursions is the most important factor in the transition rate.
\item {\it Contrary to the previous point, the Los Alamos group (Boyd and Swift 1993, unpublished but results cited in \cite{SOL95}) found that fast cooling from $T>T_c$ makes the transition easier.}  The cruicial difference is here the fast cooling started above $T_c$ in the normal phase.  This leads to smaller and more sub-domains forming during the second-order N-A transition, which allows more combinations of sub-phases collisions to be tried.
\item {\it Again contrary to the previous two points, Hakonen et al.\cite{HakKru85} found that the cooling rate has no effect on the transition rate.}  The container surface in \cite{HakKru85} is much rougher than the previous two experiments.  It leads to many sub-domains and collisions between them that the transitions happen rapidly at a catastrophic line.  It is a sign that a transition is guaranteed once the possible field excursion enters the B-phase basin of attraction, therefore independent of the cooling rate. 
\item {\it Magnetic fields affect the transitions in a complicated way not limited to its effect on $\Delta V$.}  In addition to changing the potential which affects the basin of attraction, magnetic fields also align the sub-phases of A and changes the possible field excursions to be explored.
\end{itemize}

Although these agreements are encouraging, they are only qualitative and some features can also be explained by the resonant tunneling phenomenon.  Next we will go on to make an explicit prediction that could be tested by future experiments.

 \subsection{Prediction and Discussion}
 \label{sec-pre}

The two important quantities in classical transitions are the energy threshold and the field excursions.  In reality, their temperature/pressure dependences might complicate analysis and stop us from giving a clean prediction.  However since the former is a threshold, it should not be hard to setup situations where the domain wall network is always disturbed strongly enough to allow energetic collisions.  

In this ideal situation, the transition behavior is determined only by the field excursions, and therefore the phase space structure.  As argued in Appendix~\ref{sec-pss}, it only depends on external variables through the quantity $\delta$.  Therefore, we can claim the following falsifiable prediction for our proposal:
\begin{itemize}
\item Varying only temperature and pressure, a $\delta(P,T)=const$ curve determins whether transitions are allowed or not.  
\end{itemize}
Of course, this means leaving everything else the same, including identical cooling processes through the $N-A$ transition to provide statistically similar domain wall networks. We shall leave the details of how to setup such ideal situations to experimentalists. 

There is not much room for confusion as our prediction is quite unique.  It is only shared by mechanisms which only depend on the phase space structure.  As far as we know, the only existing proposal that might have this property is related to Q-balls\cite{Hon88}, but we are not aware of a specific predictions made in that direction.  

Finally, we should point out a logical by-product of our overall argument in this paper.  It is possible for the domain wall network and classical transitions to enter the picture in a less dramatic way.  First, they provide the ``competing/parallel'' mechanism since our list in Sec.\ref{sec-exp} explains all of the (C1) to (C5) in Sec.\ref{sec-data}.  Next, we have learned two possible ways to improve the only weakness in the baked Alaska model---the quantitative prediction.

The first possibility is to keep the expanding shell conjecture as temporarily protecting the middle region, but instead of a fictitious ``fast cooling'', the middle region should be described by some kinematic rule of field excursions.   So instead of an ``energetic'' prediction which does not agree with experiments, the baked Alaska model can have a kinematic prediction similar to classical transitions.

The second possibility is to trigger domain wall collisions with incoming radiation but assume the field excursion is mostly satisfied, so the energy threshold becomes the control mechanism.  Although this is again an ``energetic'' prediction, disturbing a 2D domain wall is geometrically different from exciting a hot shell.  This can explain the lack of difference between electron and neutron radiations, and provide a more appropriate value of $n$ in the relation $\Gamma\propto-R_c^n$ to fit the data better.

Both of the above proposals require further study, and their predictions are unlikely to be as elegant as our main proposal.  Thus, at this stage, we think the easiest way to determine the significance of classical transitions is still the prediction we mentioned earlier: Transitions are controlled by the phase space structure, therefore $\delta(P,T)$.

\acknowledgments 

We thank Jason Ho, Deog Ki Hong, Piyush Kumar, Eugene Lim, Alberto Nicolis, Doug Osheroff, Jeevak Parpia, Eduardo Ponton, Grisha Volovik, Erick Weinberg and Dan Wohns for helpful discussions.  Two of our figures are taken from \cite{GibLam10} made by Tom Giblin.  This work is supported in part by the US Department of Energy and the National Science Foundation grant PHY-0855447.

\bibliographystyle{utcaps}
\bibliography{all}

\appendix

\section{The Phase Space Structure}

\label{sec-pss}

 \subsection{Direct Transitions}
Consider a general $SO(3)_s \times SO(3)_l \times U(1)$ rotation acting on the standard A-phase matrix, Eq.~(\ref{eq-A}).  The general form of all degenerate A sub-phases can be written as 
\begin{equation}
\Delta(A) = \Delta_A \vec{s}\left(\vec{u}+i\vec{v}\right)~.
\end{equation}
Here $\vec{s}$ is a real unit vector for the spin, whereas $\vec{u}$ and $\vec{v}$ are real unit external vectors orthogonal to each other, $\vec{u}\cdot\vec{v}=0$.  There are 5 degrees of freedom here.  An overall complex phase is not necessary, since it is degenerate with the rotation along $\vec{u}\times\vec{v}$.

First, we would like to find three sub-phases in A to call the left, middle, and right regions, and then apply Eq.~(\ref{eq-CTcs}) to see if the classical transition goes to the B-phase directly.  For a head-on collision, we can rotate the domain wall motion into the $x$-axis.  Then $s_\alpha u_x$ and $s_\alpha v_x$ play the role of the longitudinal mode $\psi$, while $s_\alpha u_y$, $s_\alpha v_y$, $s_\alpha u_z$ and $s_\alpha v_z$ are the transverse modes, like $\phi$.  Hence we solve the following set of equations:
\begin{eqnarray}\label{eq-directCT}
\rho(v_{Lx}\vec{s}_L+v_{Rx}\vec{s}_R)+(1-2\rho)v_{Mx}\vec{s}_M
&=&\frac{\Delta_B}{\Delta_A}\sin\theta(1,0,0)~,  \nonumber \\ 
v_{Ly}\vec{s}_L+v_{Ry}\vec{s}_R-v_{My}\vec{s}_M
&=&\frac{\Delta_B}{\Delta_A}\sin\theta(0,1,0)~,  \nonumber \\ 
v_{Lz}\vec{s}_L+v_{Rz}\vec{s}_R-v_{Mz}\vec{s}_M
&=&\frac{\Delta_B}{\Delta_A}\sin\theta(0,0,1)~,  \label{eq-CTfull} \\
\rho(u_{Lx}\vec{s}_L+u_{Rx}\vec{s}_R)+(1-2\rho)u_{Mx}\vec{s}_M
&=&\frac{\Delta_B}{\Delta_A}\cos\theta(1,0,0)~,  \nonumber \\ 
u_{Ly}\vec{s}_L+u_{Ry}\vec{s}_R-u_{My}\vec{s}_M
&=&\frac{\Delta_B}{\Delta_A}\cos\theta(0,1,0)~,  \nonumber  \\
u_{Lz}\vec{s}_L+u_{Rz}\vec{s}_R-u_{Mz}\vec{s}_M
&=&\frac{\Delta_B}{\Delta_A}\cos\theta(0,0,1)~.  \nonumber
\end{eqnarray}

Note that the left-hand sides are vector equations of spins, so we can still use $SO(3)_s$ to rotate the right-hand side into the diagonal form of the B-phase.  However the complex phase $\theta$ cannot be rotated away since the choice of collision axis already breaks $SO(3)_l$.  Counting the degrees of freedom, we have $(5\times3+1)=16$ free parameters, but $18$ equations, so the system is overconstrained.

To explicitly see that a direct transition fails, we introduce the vector
\begin{equation}
\vec{w} = \vec{v}\cos\theta - \vec{u}\sin\theta~.
\end{equation}
Since $\vec{v}\cdot\vec{u}=0$ and $\vec{v},\vec{u}\neq0$, $\vec{w}$ must be a nonzero vector.  This allows us to pair up the six equations~(\ref{eq-directCT}) for the following three:
\begin{eqnarray}
\rho(w_{Lx}\vec{s}_L+w_{Rx}\vec{s}_R)+(1-2\rho)w_{Mx}\vec{s}_M
&=&(0,0,0)~,  \\
w_{Ly}\vec{s}_L+w_{Ry}\vec{s}_R-w_{My}\vec{s}_M
&=&(0,0,0)~, \\
w_{Lz}\vec{s}_L+w_{Rz}\vec{s}_R-w_{Mz}\vec{s}_M
&=&(0,0,0)~.
\end{eqnarray}

Because the $\vec{w}$'s are nonzero vectors, it is impossible to make all three equations trivial.  At least one of them must tell us that $\vec{s}_M$ is just a linear combination of $\vec{s}_L$ and $\vec{s}_R$.  This implies that the left-hand sides of the last three equations in Eq.~(\ref{eq-directCT}) are limited to a 2D surface, while the right-hand sides are three mutually orthogonal vectors.  That is impossible.

Thus, we can conclude that classical transition cannot bring the field value directly to the bottom of the B-phase.  The next step we need to consider is the effect from its basin of attraction.

 \subsection{Basin of Attraction}
 
We need to take a closer look at the potential, Eq.~(\ref{eq-potential}) to understand the basin of attraction.  Following\cite{KauKle80}, the coefficients $\beta_i$ come from the weak coupling BCS calculation
\begin{equation}
-2\beta_1^{\rm BCS}=\beta_2^{\rm BCS}=\beta_3^{\rm BCS}
=\beta_4^{\rm BCS}=-\beta_5^{\rm BCS}~,
\end{equation}
while the strong coupling correction provides
\begin{eqnarray}
\beta_1 &=& \beta_1^{\rm BCS}(1+\frac{1}{10}\delta)~, \nonumber \\
\beta_2 &=& \beta_2^{\rm BCS}(1+\frac{1}{10}\delta)~, \nonumber \\
\beta_3 &=& \beta_3^{\rm BCS}(1-\frac{1}{40}\delta)~, \nonumber \\
\beta_4 &=& \beta_4^{\rm BCS}(1-\frac{55}{200}\delta)~, \nonumber \\
\beta_5 &=& \beta_5^{\rm BCS}(1+\frac{7}{20}\delta)~.
\end{eqnarray}
Here $\delta=\delta_{AB}=20/43$ is the critical value when $V_A=V_B$.  Smaller $\delta$ makes the B-phase more stable. 

Note that the difference between A and B phases A are only in their 4th order contraction terms.  Thus the ratio between the 2nd order and the 4th order coefficients can always be scaled away by redefining the lengh $|\Delta_{\alpha i}|$.  This will not affect the relative basin of attractions between A and B.  So the only relevant change is through $\delta$, as it affects all the 4th order coefficients differently.  We can hold everything else fixed and only vary $\delta$ to capture all physical effects relevant to the basin of attraction.

As argued in Sec.\ref{sec-boa}, the real basin of attraction is hard to delineate, but it must lie between the na\"ive and the absolute basins of attraction.

For the na\"ive basin of attraction, we perform the following numerical search.  We randomly generate three sub-phases of A to be the left, middle and right regions ($5\times3$ random variables) and combine them with the classical transition rule, Eq.~(\ref{eq-CTcs}).  Starting from this value, we follow the gradient flow $(-\nabla V)$ (an 18 dimensional vector) to see if it rolls down to the B-phase.

\begin{figure*}[ht]
\begin{center}
\includegraphics[width=8cm]{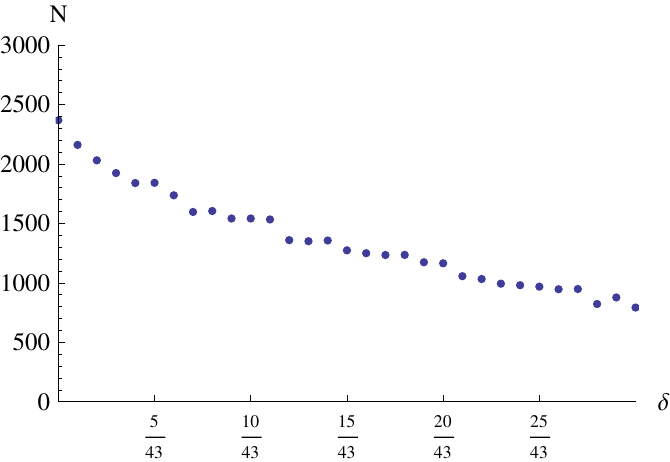}
\caption{The number of times $\Delta_{\alpha i}$ is within the B-phase na\"ive basin of attraction, as a function of $\delta$.  For each $\delta$ there are 10000 random starting points as possible values reached by collisions.  $\delta=20/43$ corresponds to $T_{AB}$ in this calculation.  The ``partial classical transition'' parameter $\rho=1/2$ in the data set here.  For $\rho=0$ and $\rho=1$, the values are roughly lower by $25\%$ and the trend remains the same.  
\label{fig-nbos}}
\end{center}
\end{figure*}

Since $0<\rho<1$ is unknown, we tried three values, $0,~1/2,~1$, and the results are similar.  As shown in Fig.\ref{fig-nbos}, decreasing $\delta$ continuously increases the chance for it to enter the B-phase.  This implies the possible value of $\Delta_{\alpha i}$ does indeed go deeper and deeper into the B-phase na\"ive basin of attraction.\footnote{It is not a relative effect that somehow the distance between the B-phase and the $15$ parameter subspace of possible collision results are getting closer.  Rather, this is a global effect that the B-phase (absolute) basin of attraction is getting larger.  We determined this by randomly choosing 10000 arbitrary $\Delta_{\alpha i}$ instead and tracking to where they roll.  The result is just Fig.\ref{fig-nbos} multiplied by a constant.}

As for the absolute basin of attraction, our approach is to first construct the $15$-dimensional compact subspace (in the $18$-dimensional full phase space of $\Delta_{\alpha i}$) that is reachable by a classical transition.  Then we locate the global minimum in this subspace.  If this restricted global minimum satisfies $V_g<V_A$, it must be in the absolute basin of attraction of phase B, since it is the only other minimum to roll to.

Again we tried three values of $\rho$ and they all have the same behavior.  For $\delta>0$, this compact subspace never intersects the B-phase absolute basin of attraction.  But zero is a critical point:  Once $\delta<0$, this compact subspace enters the absolute basin of attraction.  Note that negative $\delta$ is not a physical choice of parameter, and in fact there is a second-order phase transition here whereby the A-phase becomes a saddle point.  However we are just looking for a statement of continuity.  Since the potential $V$ and the structure of the B-phase are continuous functions of $\delta$ even through $\delta=0$, we can conclude that the B-phase absolute basin of attraction expands as $\delta$ decreases and eventually touches this $15$ dimensional subspace.

We have established that the na\"ive basin of attraction covers more and more values of $\Delta_{\alpha i}$ reached by collisions, and that the absolute basin of attraction approaches and eventually touches these values.  Therefore, the real basin of attraction, which is always between these two, must start to cover these values at some finite $\delta$, and likely covers more values at smaller $\delta$.

\end{document}